\documentclass[aps,prl,superscriptaddress,reprint,nofootinbib]{revtex4-1}

\usepackage{graphicx,multirow}
\usepackage{bm}
\usepackage{amsmath}
\usepackage{amssymb}
\usepackage{amscd}
\usepackage{latexsym}
\usepackage{slashed}
\usepackage{color}
\usepackage{graphicx}
\usepackage{hyperref}
\usepackage{ulem}
\usepackage{color}
\usepackage[caption=false]{subfig}
\usepackage{endnotes}

\def\ltap{\raisebox{-.6ex}{\rlap{$\,\sim\,$}} \raisebox{.4ex}{$\,<\,$}} 
\def\gtap{\raisebox{-.6ex}{\rlap{$\,\sim\,$}} \raisebox{.4ex}{$\,>\,$}}

\usepackage{lineno}

\begin{document}


\title{
$\mathbf{Z^\prime}$-boson dilepton searches and the high-$\mathbf{x}$ quark density
}

\author{J.~Fiaschi}%
\email{juri.fiaschi@liverpool.ac.uk}
\affiliation{Department of Mathematical Sciences, University of Liverpool, Liverpool L69 3BX}
\author{F.~Giuli}%
\email{francesco.giuli@cern.ch}
\affiliation{CERN, European Organization for Nuclear Research, 1 Esplanade des Particules, 1211 Meyrin}
\author{F.~Hautmann}%
\email{hautmann@thphys.ox.ac.uk}
\affiliation{CERN, European Organization for Nuclear Research, 1 Esplanade des Particules, 1211 Meyrin}
\affiliation{Elementaire Deeltjes Fysica, Universiteit Antwerpen, B 2020 Antwerpen}
\affiliation{Theoretical Physics Department, University of Oxford, Oxford OX1 3PU}%
\author{S.~Moch}
\email{sven-olaf.moch@desy.de}
\affiliation{II. Institut f{\" u}r Theoretische Physik, Universit{\" a}t Hamburg, Luruper Chaussee 149, 22761 Hamburg}
\author{S.~Moretti}
\email{s.moretti@soton.ac.uk; stefano.moretti@physics.uu.se}
\affiliation{School of Physics and Astronomy, University of Southampton, Highfield, Southampton SO17 1BJ}
\affiliation{Department of Physics and Astronomy, Uppsala University, Box 516, SE-751 20 Uppsala}

\begin{abstract}
\noindent
We study the influence of theoretical systematic uncertainties due to the quark density on LHC experimental searches for $Z^\prime$-bosons.
Using an approach originally proposed in the context of the ABMP16 PDF set for the high-$x$ behaviour of the quark density, we presents results on observables commonly used to study $Z^\prime$ signals in dilepton channels.
\end{abstract}

\vspace*{-0.8cm} 

\hspace*{4.5cm} CERN-TH-2022-177

\vspace*{0.2cm}

\hspace*{6.1cm}    LTH 1323

\vspace*{0.2cm}

\hspace*{5.55cm}    DESY-22-178

\vspace*{0.3cm}

\maketitle

{\it Introduction.} 
Beyond-the-Standard-Model (BSM) searches for heavy neutral spin-1 $Z^\prime$-bosons~\cite{Langacker:2008yv} are carried out at the Large Hadron Collider (LHC)~\cite{Salvioni:2009mt,Accomando:2010fz} by the ATLAS~\cite{ATLAS:2019erb} and CMS~\cite{CMS:2021ctt} collaborations and will continue at the High-Luminosity LHC (HL-LHC)~\cite{CidVidal:2018eel,Gianotti:2002xx} and future colliders~\cite{Golling:2016gvc}.
For narrow $Z^\prime$ states, experimental analyses rely on resonant mass searches.
For BSM scenarios involving $Z^\prime$ states with large width, the collider sensitivity to new physics signals is influenced significantly by the theoretical 
modelling of both signal and background~\cite{Accomando:2019ahs}. 
In this case, one of the main sources of systematic uncertainties, affecting the potential of experimental searches for discovering or setting exclusion bounds on new $Z^\prime$ bosons, is given by the Parton Density Functions (PDFs), describing the Quantum Chromo-Dynamics (QCD) evolution of the initial partonic states entering the collision. 

The crucial role of the PDFs can be illustrated, e.g., by the analysis~\cite{Fiaschi:2021sin} of Drell-Yan (DY) processes, i.e., dilepton channels in hadronic collisions. 
This paper examines broad vector resonances in BSM strongly-interacting Higgs models~\cite{Panico:2015jxa,Giudice:2007fh}, using the 4-Dimensional Composite Higgs Model (4DCHM) realisation~\cite{DeCurtis:2011yx} of the minimal composite Higgs scenario~\cite{Agashe:2004rs}.
The PDF systematic uncertainties are dominated by the quark sector, for moderate to large momentum fractions $x$.
It is shown in Ref.~\cite{Fiaschi:2021okg}, by an \texttt{xFitter}~\cite{Alekhin:2014irh,xFitter:2022zjb} ``profiling'' analysis, that the quark PDF systematics in this region can be improved by exploiting, besides unpolarised DY production, the experimental information on the SM vector boson polarisation in the DY mass range near the SM vector boson peak.
More precisely, ``improved PDFs''~\cite{Fiaschi:2021okg} are obtained by combining 
high-statistics precision measurements of DY lepton-charge and forward-backward~\cite{Accomando:2019vqt,Abdolmaleki:2019ubu,Accomando:2018nig,Accomando:2017scx} asymmetries, associated with the difference between the left-handed and right-handed vector boson polarisation fractions. 
Ref.~\cite{Fiaschi:2021sin} demonstrates that the LHC sensitivity to the BSM large-width states is then greatly increased compared to the case of base PDF sets. 
 
The purpose of this note is to investigate further aspects of the PDF systematics in the 
$Z^\prime$-boson dilepton search region, concentrating on the effect, pointed out in 
Ref.~\cite{Alekhin:2017kpj}, of the $x \to 1$ behaviour of the quark densities.
The very high mass tails of physical distributions are influenced by the quark density at 
high $x$ and low mass scales through QCD evolution, in particular through a combination of non-diagonal contributions to the flavour-singlet evolution 
kernel~\cite{Alekhin:2017kpj}.
This sensitivity can be recast as a parameterisation uncertainty $(1-x)^b$ in the nearly-vanishing quark density as $x \to 1$~\cite{Courtoy:2020fex}.
We will follow the approach proposed in~\cite{Alekhin:2017kpj} to treat the $x \to 1$ quark density and analyse its impact on the multi-TeV mass region relevant for BSM $Z^\prime$ searches. 

We remark that this approach can be regarded as a method to take into account certain kinds of 
theoretical uncertainties, associated with PDF determinations, which go beyond the standard uncertainties provided by the main PDF sets.
Full approaches to producing PDF sets with theory uncertainties are currently being investigated by several groups, e.g., see~\cite{NNPDF:2019ubu,McGowan:2022nag} and ongoing \texttt{xFitter}~\cite{xFitter:2022zjb} studies based on the method of~\cite{Bertone:2022ope,Bertone:2022sso}. 
For the problem of interest in this paper 
we use the simple approach of~\cite{Alekhin:2017kpj}, as a complete study will only become possible once PDF sets with full theory uncertainties are available.

We proceed in the following manner.
We start by describing the criteria to construct the PDF ensemble through which we parameterise systematic uncertainties in the quark density at high $x$.
Then we consider dilepton observables which are commonly used for BSM searches in the multi-TeV region, the dilepton invariant mass spectrum 
of the differential cross section, $ d \sigma / d M_{\ell \ell} $,  and reconstructed 
forward-backward asymmetry $A_{\rm{FB}}^*$. The reconstruction follows the standard prescription based on the boost of the dilepton system, as described in 
Ref.~\cite{Accomando:2015cfa}. We study these observables in two stages. 

First, we compute $ d \sigma / d M_{\ell \ell} $ and $A_{\rm{FB}}^*$ in the SM by including the high-$x$ quark density systematics.
We compare these results both with those based on the original ABMP16 set~\cite{Alekhin:2017kpj} from which the PDF ensemble to study the high-$x$ systematics is derived and with those from other commonly used PDF sets, CT18~\cite{Hou:2019efy}, MSHT20~\cite{Bailey:2020ooq} and NNPDF4.0~\cite{NNPDF:2021njg}.
This provides us with a validation test and a quantitative estimate of the impact of high-$x$ systematics on SM predictions.
An  investigation of the asymmetry $A_{\rm{FB}}^*$ in the SM has been recently carried 
out in~\cite{Ball:2022qtp}.
Further studies on the $A_{\rm{FB}}^*$ in the context of PDF determinations may be found 
in~\cite{Yang:2022zvx,Yang:2022bxv,Xie:2022tzo,Yang:2021cpd,Fu:2020mxl,Willis:2018yln,Accomando:2016tah,Accomando:2016ehi,Bodek:2015ljm}.

Second, we consider BSM benchmark models for heavy $Z^\prime$ bosons representative of the $E_6$, Generalised Left-Right (GLR) and Generalised Standard Model (GSM)~\cite{Accomando:2010fz}.
We study the signal profiles in the invariant mass spectrum of the cross section and $A_{\rm{FB}}^*$, by discussing them in the light of the high-$x$ quark density systematics.

\begin{figure}[t!]
\begin{center}
\includegraphics[width=0.40\textwidth]{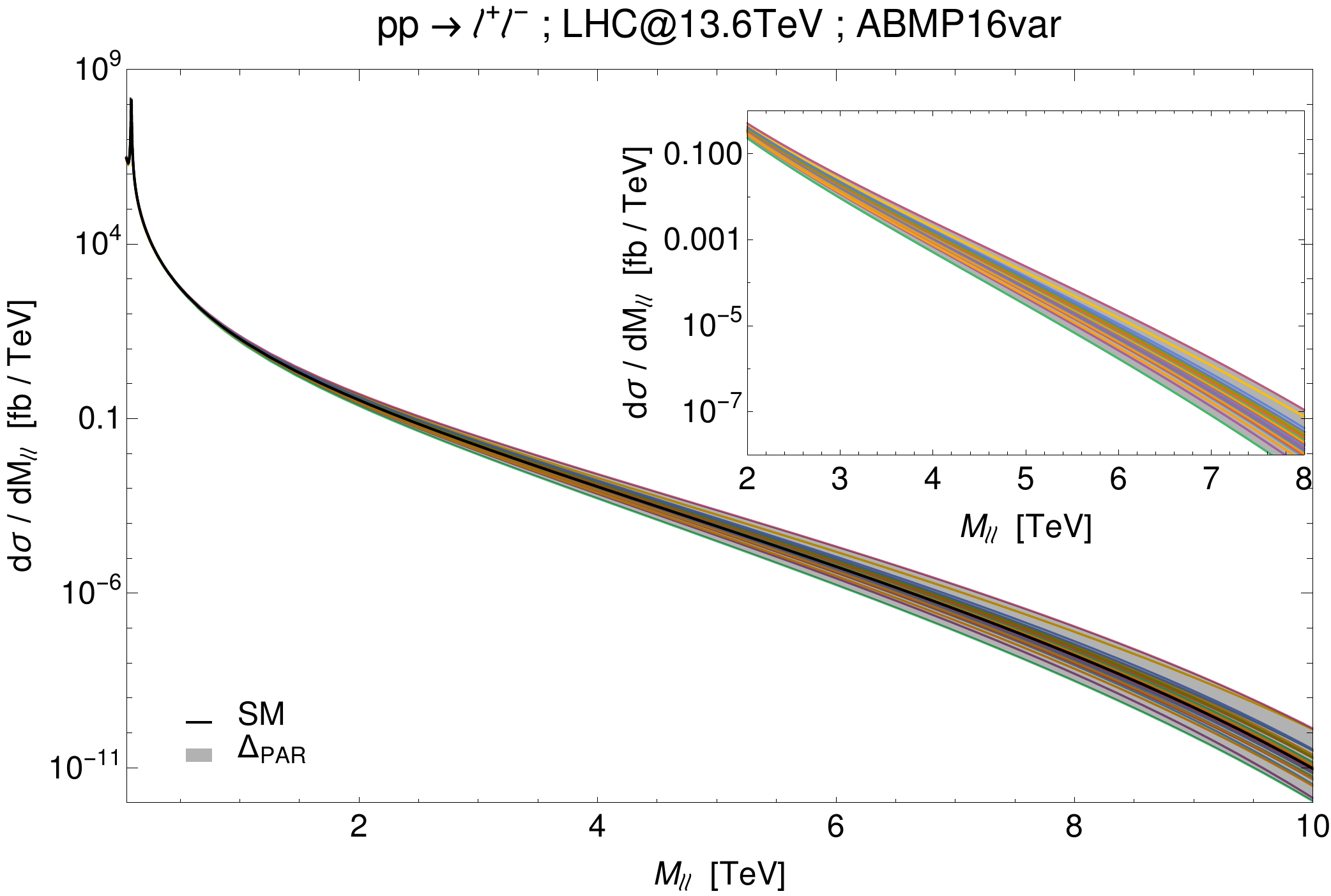}
\includegraphics[width=0.40\textwidth]{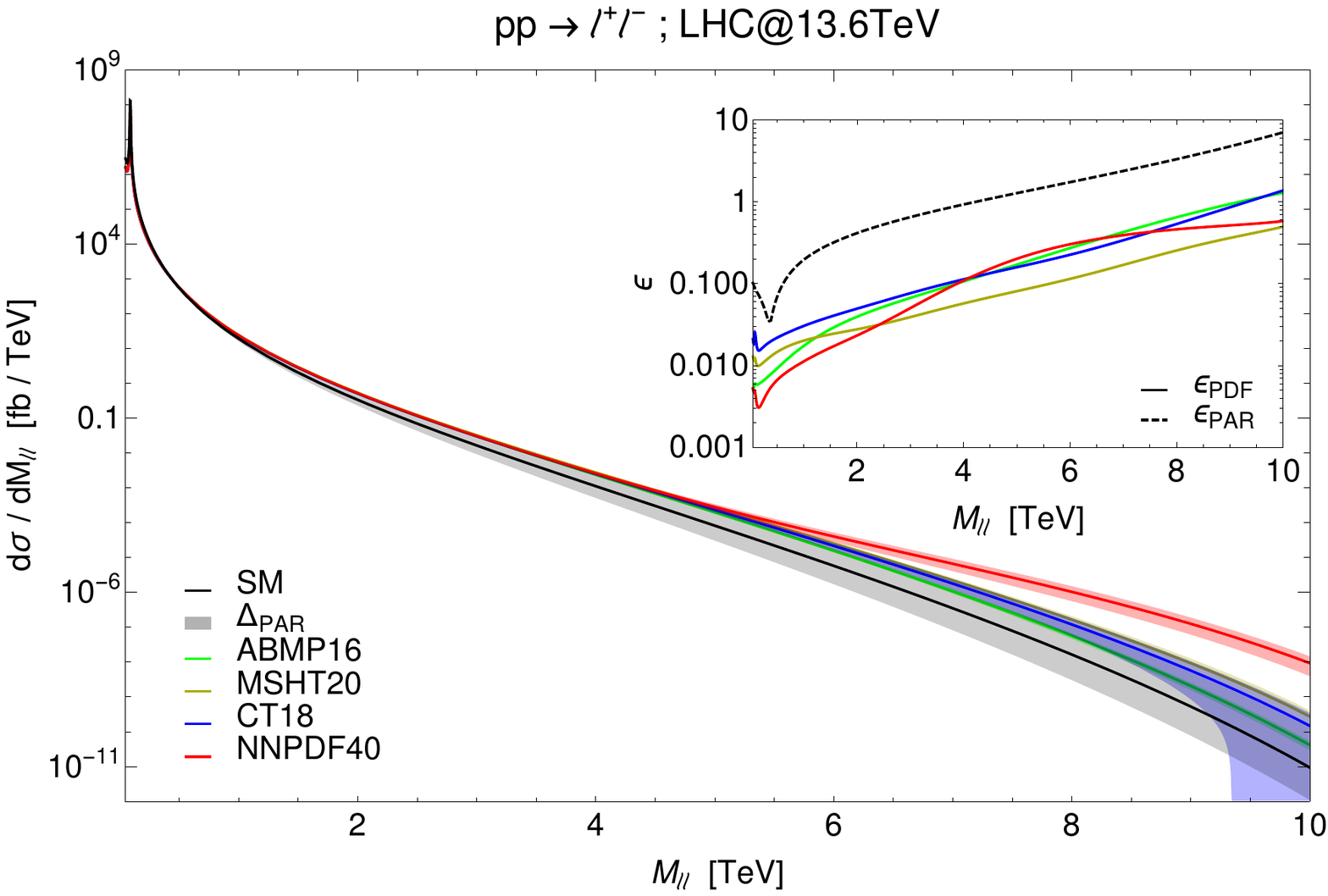}
\caption{Differential cross section in invariant mass of the dilepton final state at the LHC with $\sqrt{s}$ = 13.6 TeV:  
 (top) results for all ABMP16var members, with the inset plot 
 focusing on the few TeV invariant mass region; (bottom) ABMP16var result 
 compared with results from standard PDF sets,  
 with the inset plot showing the relative sizes of parameterisation and ``standard" PDF uncertainties.
}
\label{fig:invmass}
\end{center}
\end{figure}

{\it Quark density at large $x$.} 
To study the impact of high-$x$ quark densities on 
dilepton observables, we construct a PDF ensemble 
as follows. 
The ``central'' value of the ensemble is obtained using the results of the fit contained in Tab.VII of Ref.~\cite{Alekhin:2017kpj}.
We then construct 24 ensemble members by varying the exponent of the $(1 - x)$ term of the parameterisation of $u$ and $d$ quarks and antiquarks by $\pm$ 0.3, 0.5, 1.0.
We obtain 
 predictions for dilepton observables 
using 22 members of the set, 
excluding the members with 
variations by $+1$ for $u$ and by $-1$ for $d$ as these do not 
yield a sufficiently fast fall-off in the ratio $d / u$ for $x \to 1$. 
We further include predictions obtained from the following additional 
members. 

1. All quark and antiquark distributions varied simultaneously by $\pm$ 0.3, 0.5, 1 (i.e., 6 additional variations). We name this Variation \#1.

2. Quark distributions varied by $\pm$ 1  and antiquark distributions 
varied by $\mp$ 1  simultaneously (i.e., 2 additional variations).
We name this Variation \#2.

All these combinations ensure a vanishing $d/u$ ratio as $x \to 1$.
In the following the predictions obtained from the PDF ensemble 
described above will be referred to as ``ABMP16var'', and the envelope from the 30 different variations will be used as an estimate of the parameterisation uncertainty.
We will investigate the impact of this additional uncertainty on observables of the dilepton final state.

\begin{figure}[t!]
\begin{center}
\includegraphics[width=0.40\textwidth]{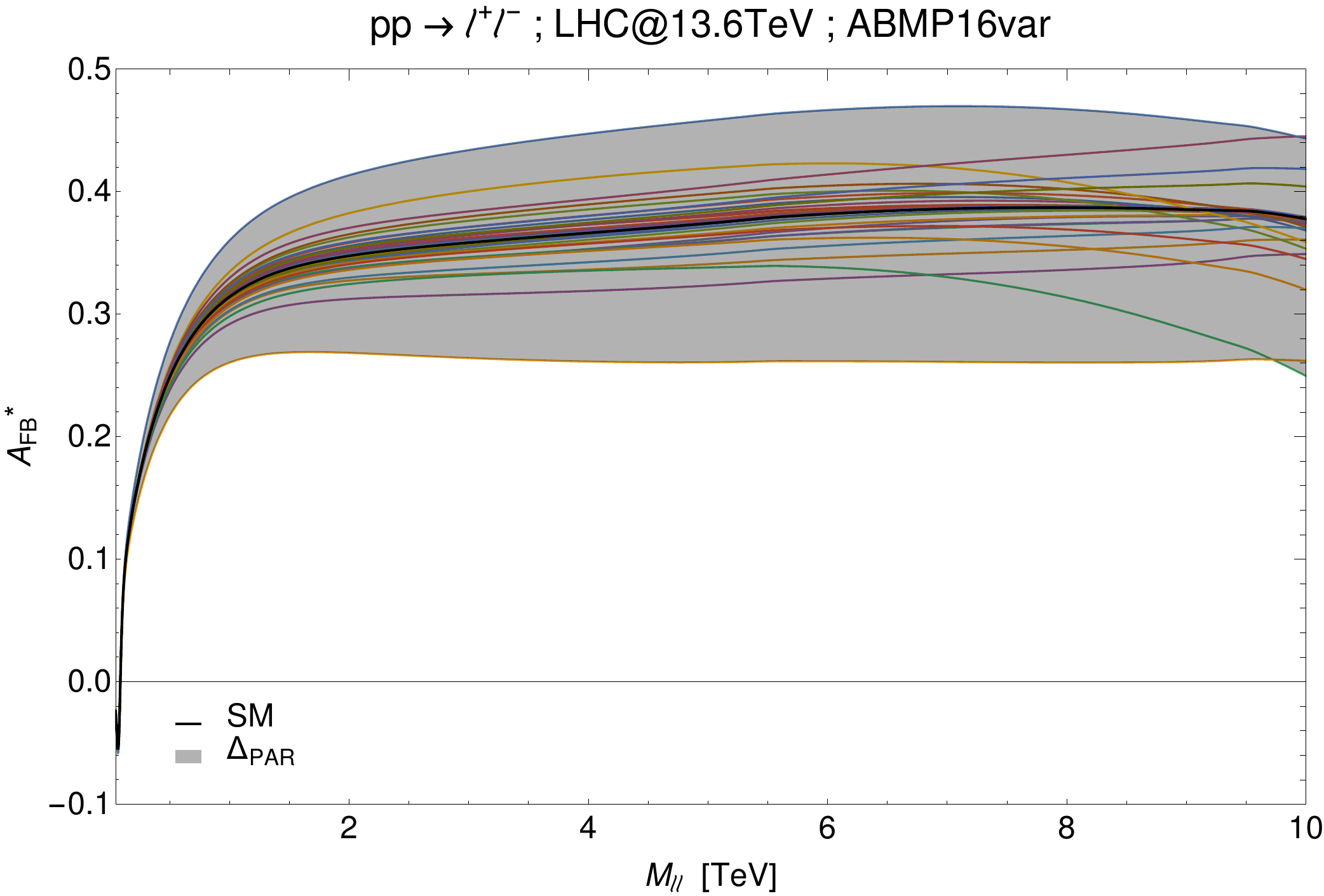}
\includegraphics[width=0.40\textwidth]{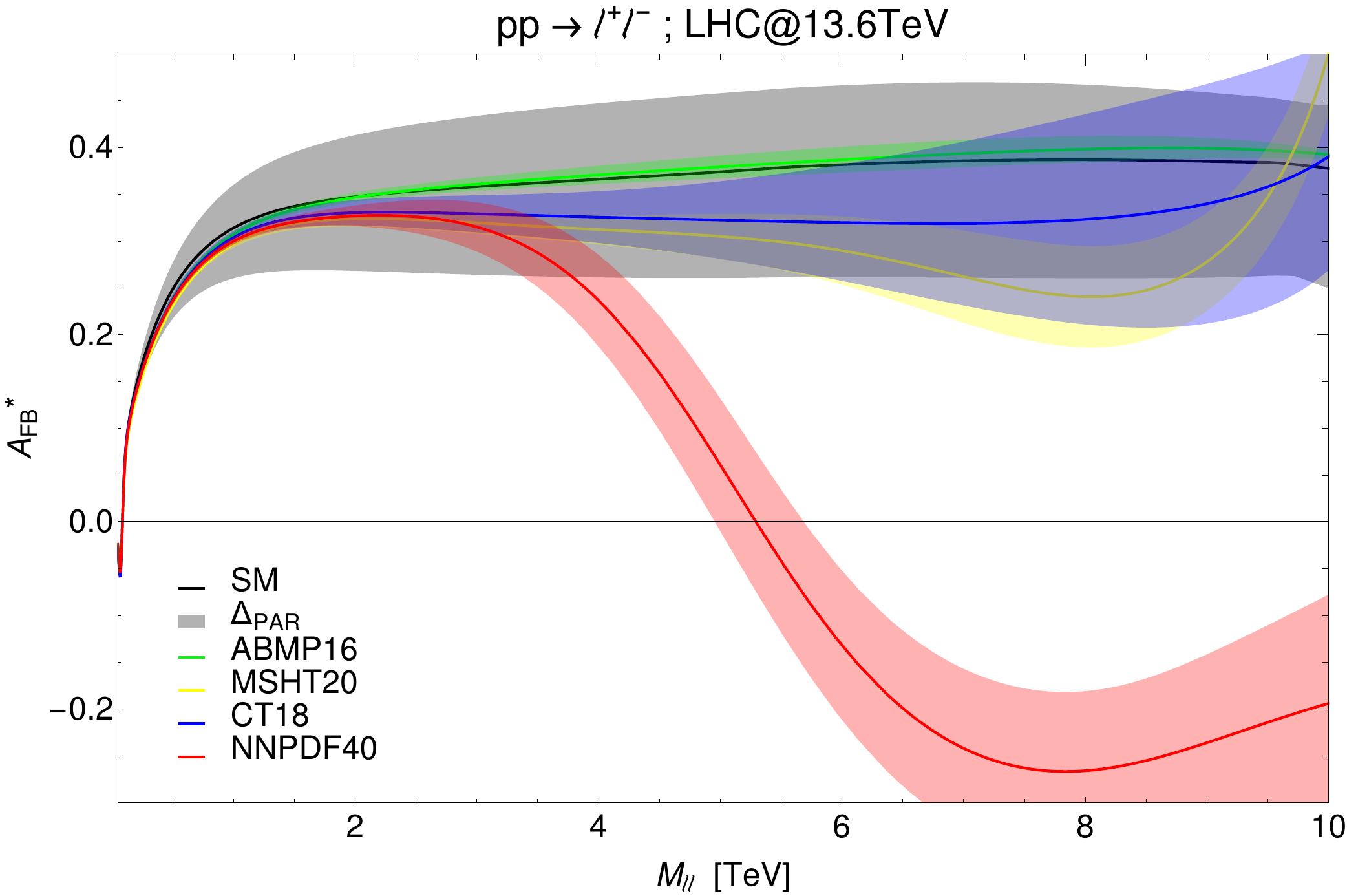}
\caption{Asymmetry $A_{\rm{FB}}^*$ in 
invariant mass of the dilepton final state at the LHC with $\sqrt{s}$ = 13.6 TeV: 
 (top) results for all ABMP16var members; (bottom) 
 ABMP16var result compared with results from 
 standard PDF sets.}
\label{fig:afb-SM}
\end{center}
\end{figure}

{\it SM results.} Using the above method ABMP16var, 
we now compute the dilepton invariant mass distribution of the 
differential cross section $ d \sigma / d M_{\ell \ell} $  and 
forward-backward asymmetry $A_{\rm{FB}}^*$ in the SM. The results 
are given  in  Fig.~\ref{fig:invmass} for $ d \sigma / d M_{\ell \ell} $ and in 
Fig.~\ref{fig:afb-SM} for $A_{\rm{FB}}^*$. 

The top panels of these figures    
show the results for all the variations in the ABMP16var ensemble, and the 
parameterisation uncertainty $\Delta_{\rm{PAR}}$ from their envelope.  
 For  $ d \sigma / d M_{\ell \ell} $ the 
largest variations are obtained from  Variation \#1, while for 
$A_{\rm{FB}}^*$ the largest variations are obtained from  Variation \#2. 
The inset plot of the top panel in  Fig.~\ref{fig:invmass} 
focuses on the few TeV invariant mass region, relevant for BSM searches at the LHC.

In the bottom panels of Figs.~\ref{fig:invmass}  and~\ref{fig:afb-SM},  
the ABMP16var results are compared with the results from 
the original ABMP16 set~\cite{Alekhin:2017kpj} and  
 other commonly used sets 
CT18~\cite{Hou:2019efy}, MSHT20~\cite{Bailey:2020ooq} and 
NNPDF4.0~\cite{NNPDF:2021njg}.  The 
 parameterisation uncertainty $\Delta_{\rm{PAR}}$ is compared with 
 the ``standard'' PDF uncertainties for each set. 
 The bottom panel of Fig.~\ref{fig:invmass} 
 indicates that 
  the predictions for the SM differential cross section 
 agree for  all PDF sets  within uncertainties, except  
 NNPDF4.0 which departs from the others for $ M_{\ell \ell} \gtap  5$ TeV. 
 The inset plot of the bottom panel in  Fig.~\ref{fig:invmass} displays the 
 relative sizes $\epsilon = \Delta\sigma / \sigma$ of the parameterisation 
 uncertainties ($\epsilon_{\rm{PAR}}$) and   PDF uncertainties of 
 each set ($\epsilon_{\rm{PDF}}$), illustrating  
 that the former is roughly one order of magnitude larger than the latter.

The bottom panel of Fig.~\ref{fig:afb-SM}  gives the SM predictions 
for the forward-backward asymmetry  $A_{\rm{FB}}^*$ from different  PDF sets. 
It shows that  the  
NNPDF4.0~\cite{NNPDF:2021njg}   set  
leads to  results  for the  $A_{\rm{FB}}^*$  which 
are similar to those of the 
ABMP16~\cite{Alekhin:2017kpj}, 
CT18~\cite{Hou:2019efy} and MSHT20~\cite{Bailey:2020ooq}  sets 
for $ M_{\ell \ell} \ltap 4$ TeV but, for  $ M_{\ell \ell} \gtap  4$ TeV, start to differ 
dramatically from those of the other sets, which remain similar, within 
PDF  uncertainties,  among each other. 
The NNPDF prediction acquires a negative slope, decreases toward zero and  
becomes negative above  $ M_{\ell \ell}   \approx 5$ TeV,   while the 
 predictions from the other sets stay positive,  and roughly flat. 
The NNPDF behaviour 
has been discussed at length in the recent work of 
Ref.~\cite{Ball:2022qtp}. Here 
we limit ourselves to making a few general remarks, particularly in 
the light of the method of the present paper 
to study the high-$x$ quark density systematics.

The peculiar   NNPDF   
prediction for   $A_{\rm{FB}}^*$    at large masses in the SM 
 stems from the fact that the 
NNPDF4.0~\cite{NNPDF:2021njg}  set has antiquark distributions falling 
 much more slowly  at high $x$ than all the other sets and, conversely,  
 quark distributions falling much faster at high $x$ than the other 
 sets~\cite{Ball:2022qtp}. 
 This ``high-sea, low-valence" scenario 
 applies to all light flavours, and is especially pronounced 
for the ${\overline u}$ and $d$ distributions.   In general, for   $A_{\rm{FB}}^*$ 
to stay positive one has to have the slope of the antiquark's  fall-off 
to be larger than the slope of the quark's fall-off for all values of $x$. 
 This condition is fulfilled by the PDF 
sets~\cite{Alekhin:2017kpj,Hou:2019efy,Bailey:2020ooq}, but it is 
not fulfilled by the NNPDF set~\cite{NNPDF:2021njg}, starting approximately 
for $x \gtap 0.4$~\cite{Ball:2022qtp}. This 
feature of the NNPDF sea and valence quark distributions,  
 particularly in the region $ 0.4 \ltap x \ltap 0.6$, is responsible for    
 the NNPDF  $A_{\rm{FB}}^*$  prediction in the bottom panel of 
 Fig.~\ref{fig:afb-SM} 
 decreasing and turning negative 
around $ M_{\ell \ell}   \approx 5$ TeV.

\begin{figure}[t!]
\begin{center}
\includegraphics[width=0.50\textwidth]{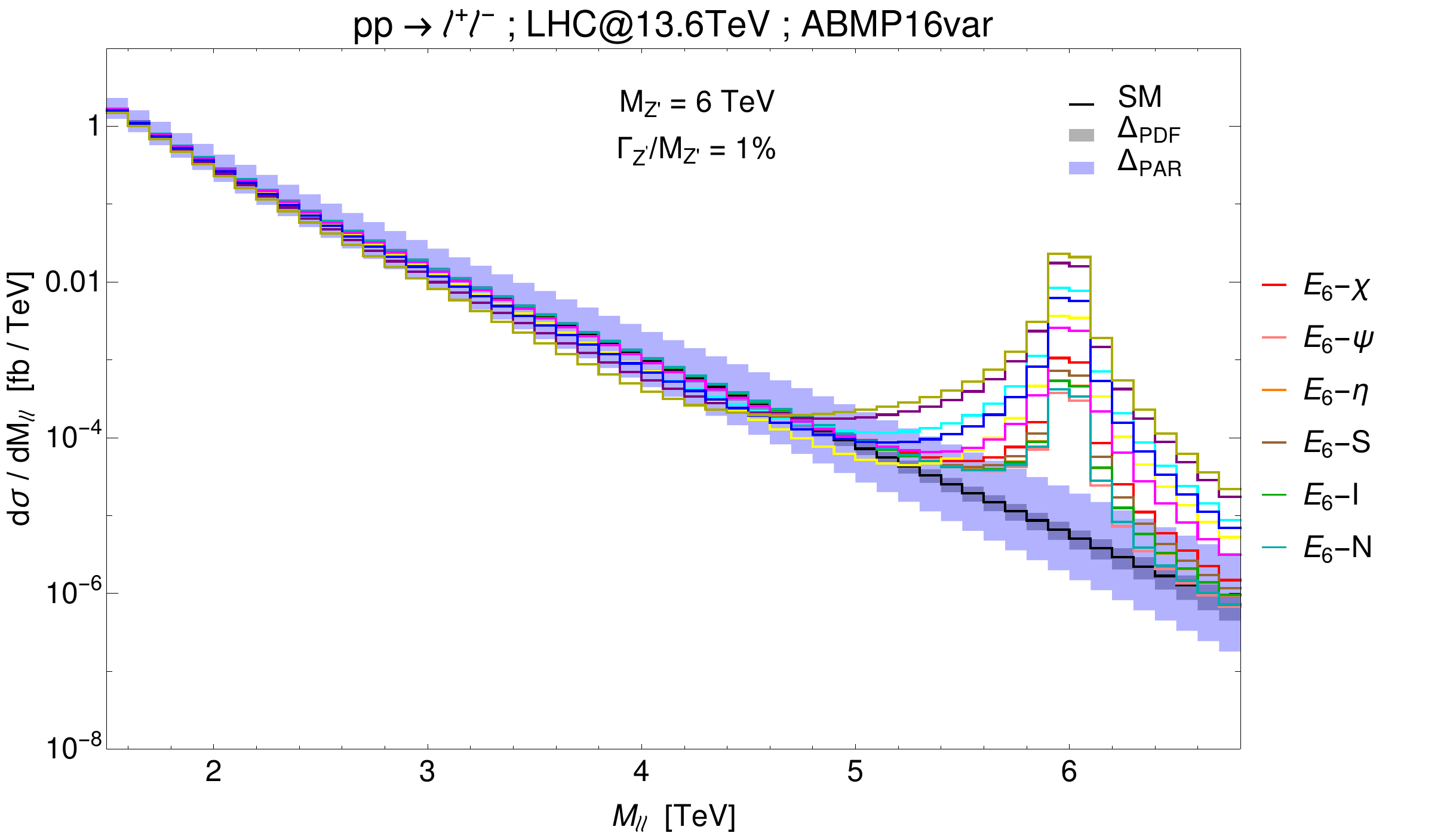}
\includegraphics[width=0.50\textwidth]{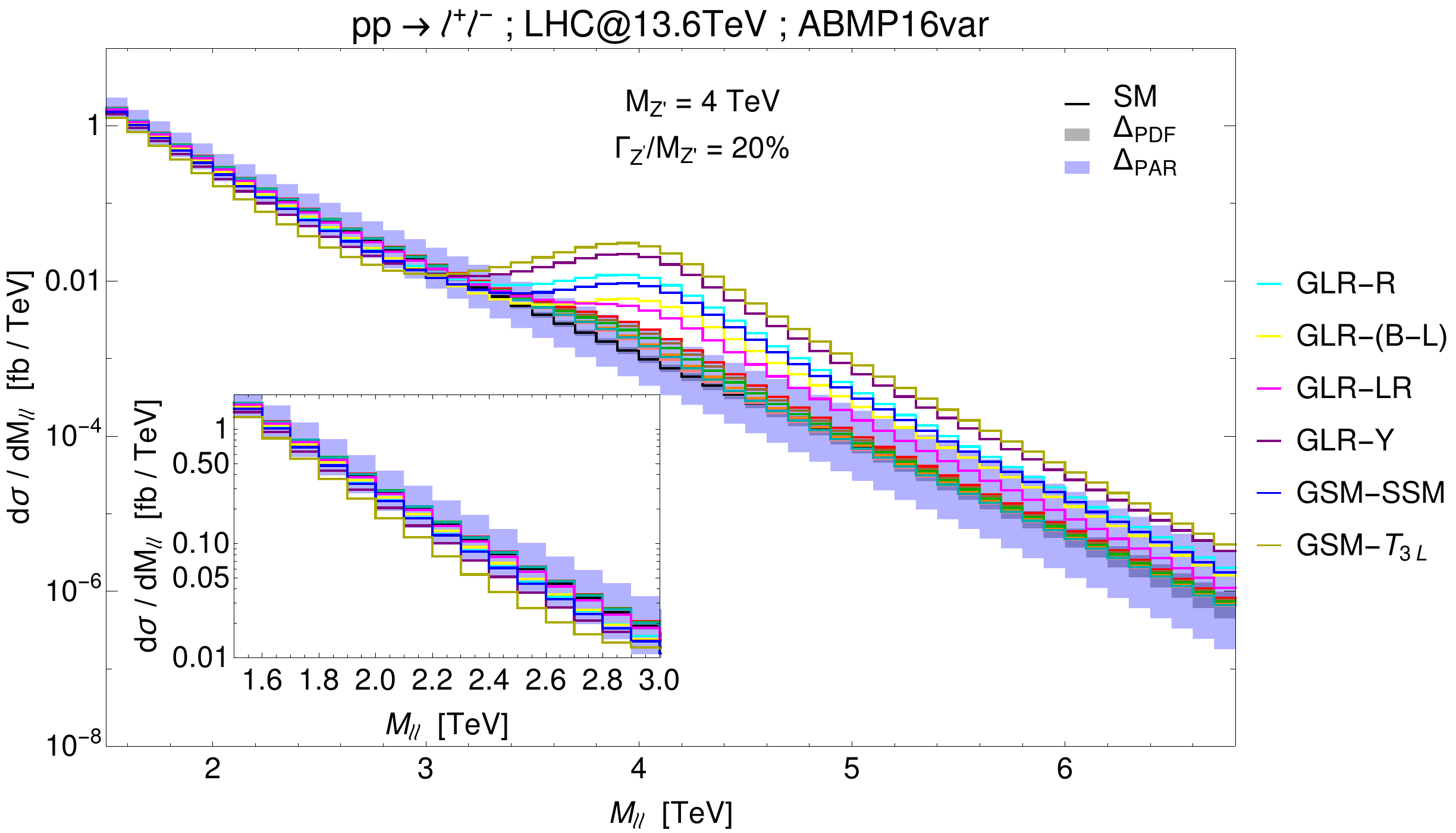}
\caption{Differential cross section in invariant mass of the dilepton final state at the LHC with $\sqrt{s} = 13.6$ TeV.
Coloured curves are obtained for a series of single $Z^\prime$ benchmark models, where we set (top) $M_{Z^\prime} = 6$ TeV, $\Gamma_{Z^\prime} / M_{Z^\prime} = $ 1\% and (bottom) $M_{Z^\prime} = 4$ TeV, $\Gamma_{Z^\prime} / M_{Z^\prime} = $ 20\%.}
\label{fig:bsm-invmass}
\end{center}
\end{figure}

In order to constrain quark distributions in the 
 region $ 0.4 \ltap x \ltap 0.6$,  fixed-target deep inelastic 
 scattering and forward Drell-Yan production data are usually employed. For 
 instance, in the 
 case of the ABMP16 set~\cite{Alekhin:2017kpj} it is primarily the 
 forward  LHCb~\cite{LHCb:2015mad,LHCb:2015okr,LHCb:2015kwa} 
 and  D0~\cite{D0:2014kma}  
 data which determine the quark distributions in this region, and lead to 
  low enough sea distributions, compared to the valence distributions,  
that the SM prediction for $A_{\rm{FB}}^*$ stays positive.   
  To fully assess the high-mass asymmetry SM predictions in  
  Fig.~\ref{fig:afb-SM},    it will be relevant both a) to 
  verify the compatibility of  high-sea  NNPDF-like  scenarios 
  with   forward Drell-Yan production from present and upcoming 
  measurements, and b) to estimate the impact 
  of  large-$x$  systematic uncertainties, such as those 
  taken into account via the  ABMP16var method,   
  on  high-sea NNPDF-like  scenarios. 
  We see from  Fig.~\ref{fig:afb-SM}  that, in the 
  low-sea ABMP-like case,  these large-$x$ 
  systematic uncertainties, although 
 significantly larger than the  ``standard'' PDF uncertainties, do not however 
  endanger the  $A_{\rm{FB}}^*$ positivity in the SM. On the other hand, it cannot 
be  ruled out  that, in 
the high-sea NNPDF-like case, such large-$x$ systematic 
  uncertainties might yield an error band on the 
  $A_{\rm{FB}}^*$ result in the high-mass region so large as  
  to overcome   the difference currently seen between predictions from 
  different PDF sets. Investigations of this are warranted. 
  This possibility  underlines the relevance, even within the SM, of the 
  studies of  high-$x$ quark density systematics proposed in the present work.

\begin{figure}[t!]
\begin{center}
\includegraphics[width=0.50\textwidth]{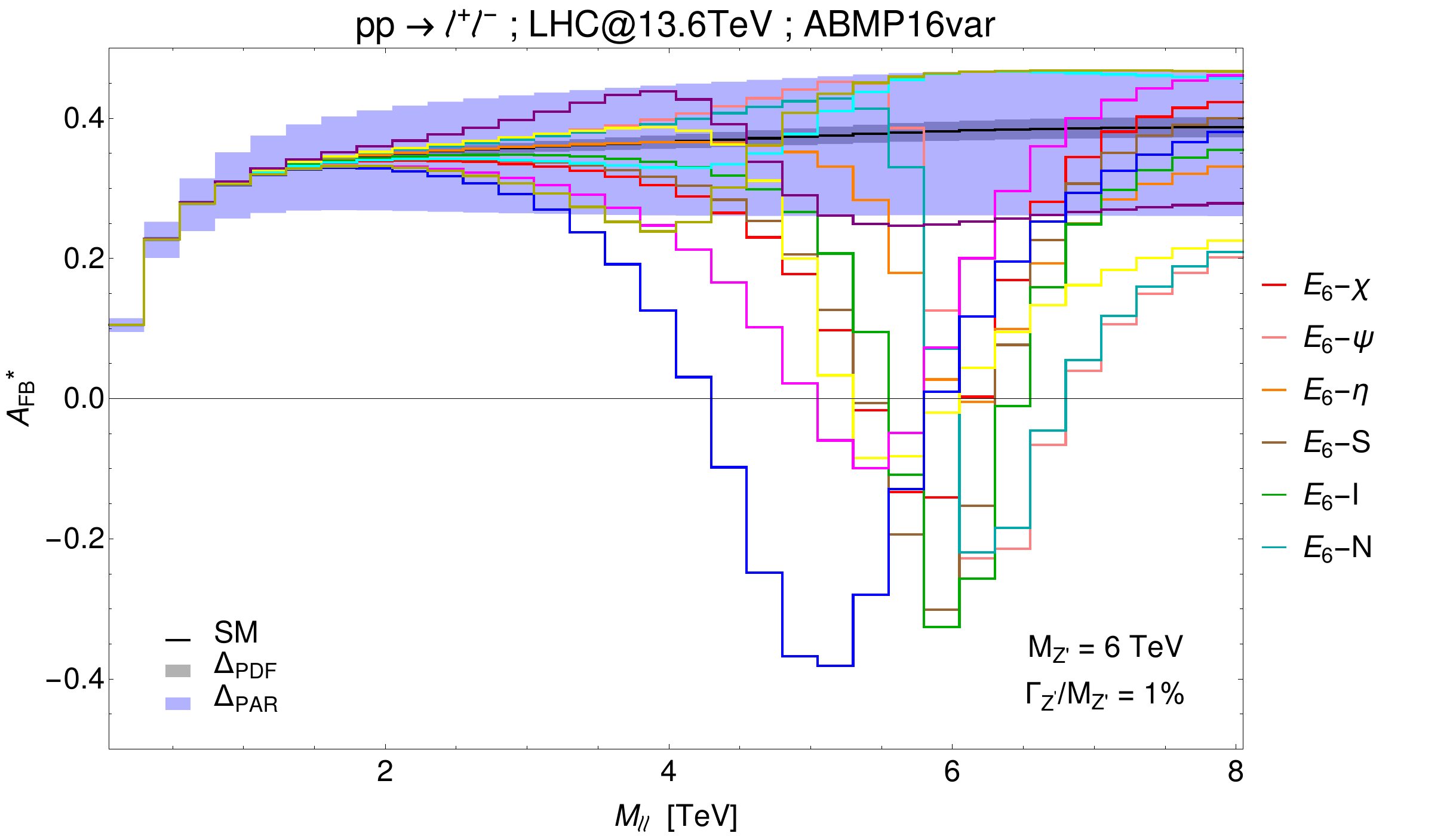}
\includegraphics[width=0.50\textwidth]{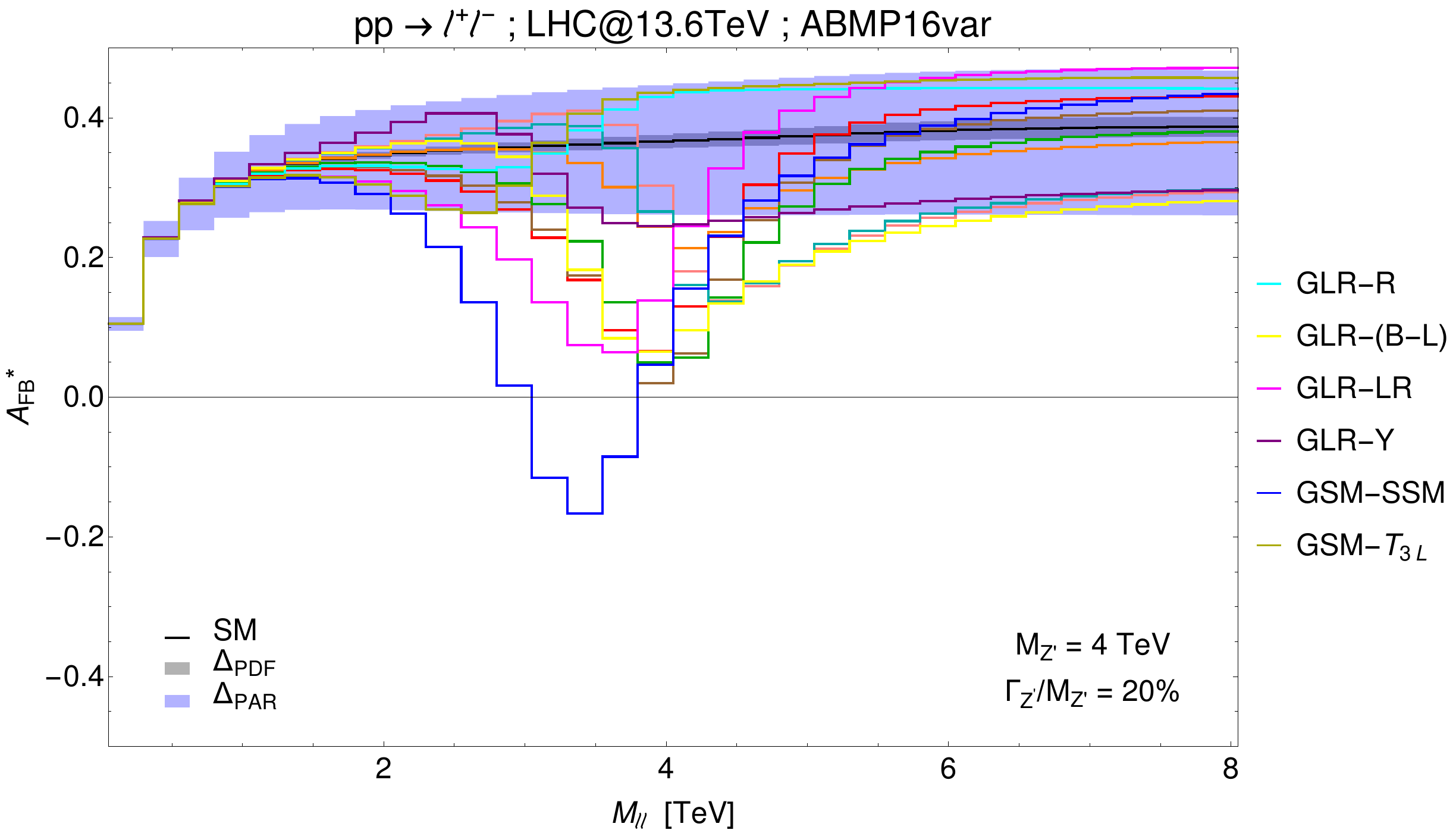}
\caption{Asymmetry 
$A_{\rm{FB}}^*$ in invariant mass of the dilepton final state 
at the LHC with $\sqrt{s} = 13.6$ TeV.
Coloured curves are obtained for a series of single $Z^\prime$ benchmark models, where we set (top) $M_{Z^\prime} = 6$ TeV, $\Gamma_{Z^\prime} / M_{Z^\prime} = $ 1\% and (bottom) $M_{Z^\prime} = 4$ TeV, $\Gamma_{Z^\prime} / M_{Z^\prime} = $ 20\%.
}
\label{fig:afb-bsm}
\end{center}
\end{figure}

{\it BSM searches.} 
As the systematic uncertainty from the high-$x$ quark density is generally comparable to or larger than the ``standard'' PDF error, it is important to investigate it in the context of BSM searches.
In Fig.~\ref{fig:bsm-invmass} we show the signal profile in the differential cross section observable for narrow (top) and wide (bottom) $Z^\prime$-bosons from a series of benchmark models.
They belong to Grand Unified Theory (GUT) inspired classes of models predicting a naturally heavy (narrow) $Z^\prime$ and have been described in~\cite{Accomando:2010fz}.
While narrow resonances with masses below 5 TeV are excluded by direct searches from ATLAS~\cite{ATLAS:2019erb} and CMS~\cite{CMS:2021ctt}, lighter wide resonances as heavy as few TeV are still allowed, depending on their specific realisation.
We set the mass of narrow $Z^\prime$s ($\Gamma_{Z^\prime} / M_{Z^\prime} = 1\%$) at 6 TeV and of wide $Z^\prime$s ($\Gamma_{Z^\prime} / M_{Z^\prime} = 20\%$) at 4 TeV, so as to comply with current bounds.
Narrow models would be marginally affected by the additional source of uncertainty, while broad resonances would potentially suffer a strong reduction of sensitivity.
An interesting feature appearing more visibly in the case of wide resonances concerns the negative interference contribution occurring in the low mass tail of the distribution.
The resulting depletion of events, while potentially affecting the control region in the analysis for BSM searches, can also lead to an early indication of the presence of a 
BSM contribution~\cite{Accomando:2019ahs, Fiaschi:2021sin}, above and beyond SM uncertainties (for some $Z^\prime$ scenarios), see inset in the bottom plot.

\begin{figure}[t!]
\begin{center}
\includegraphics[width=0.40\textwidth]{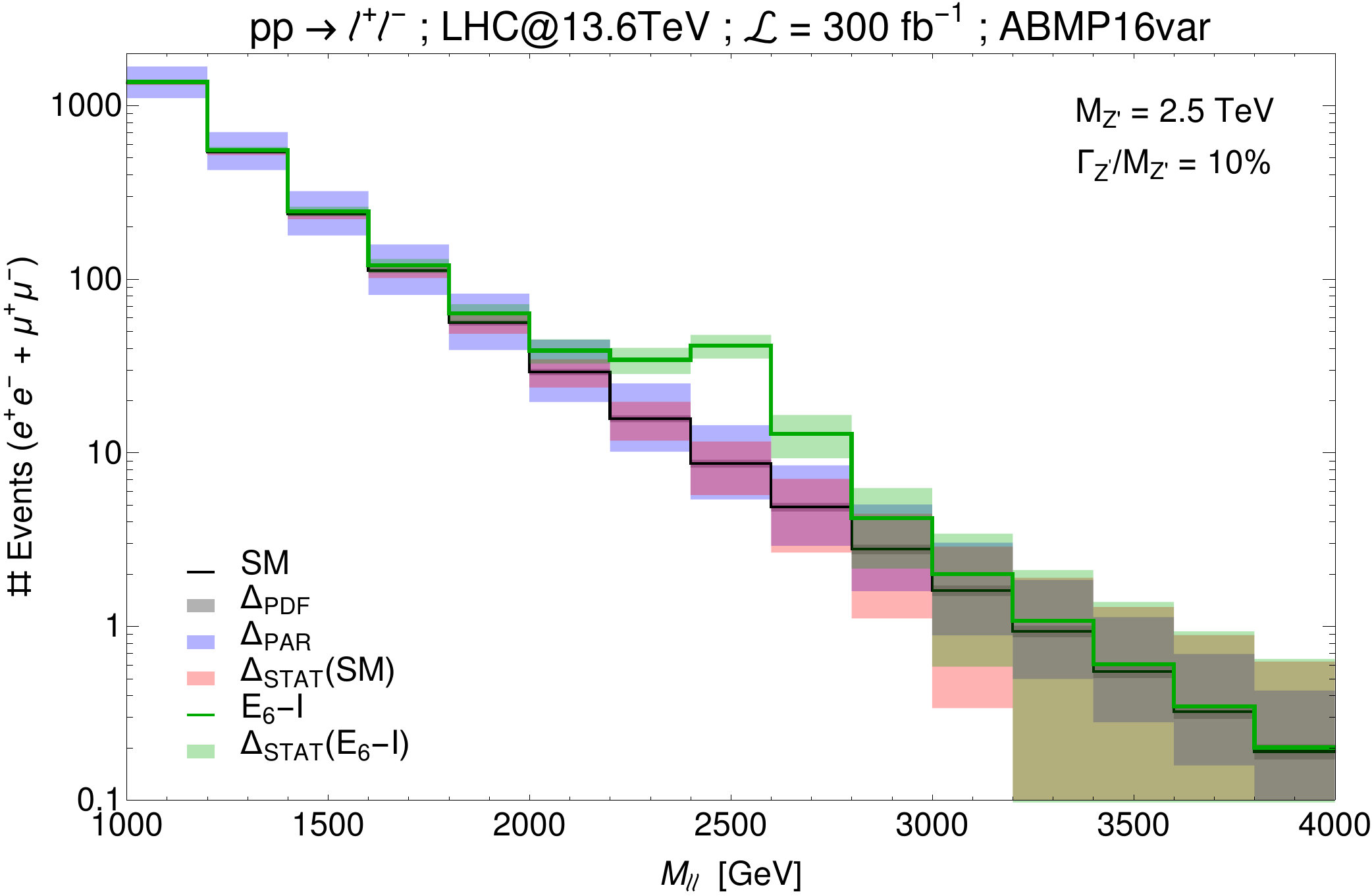}
\includegraphics[width=0.40\textwidth]{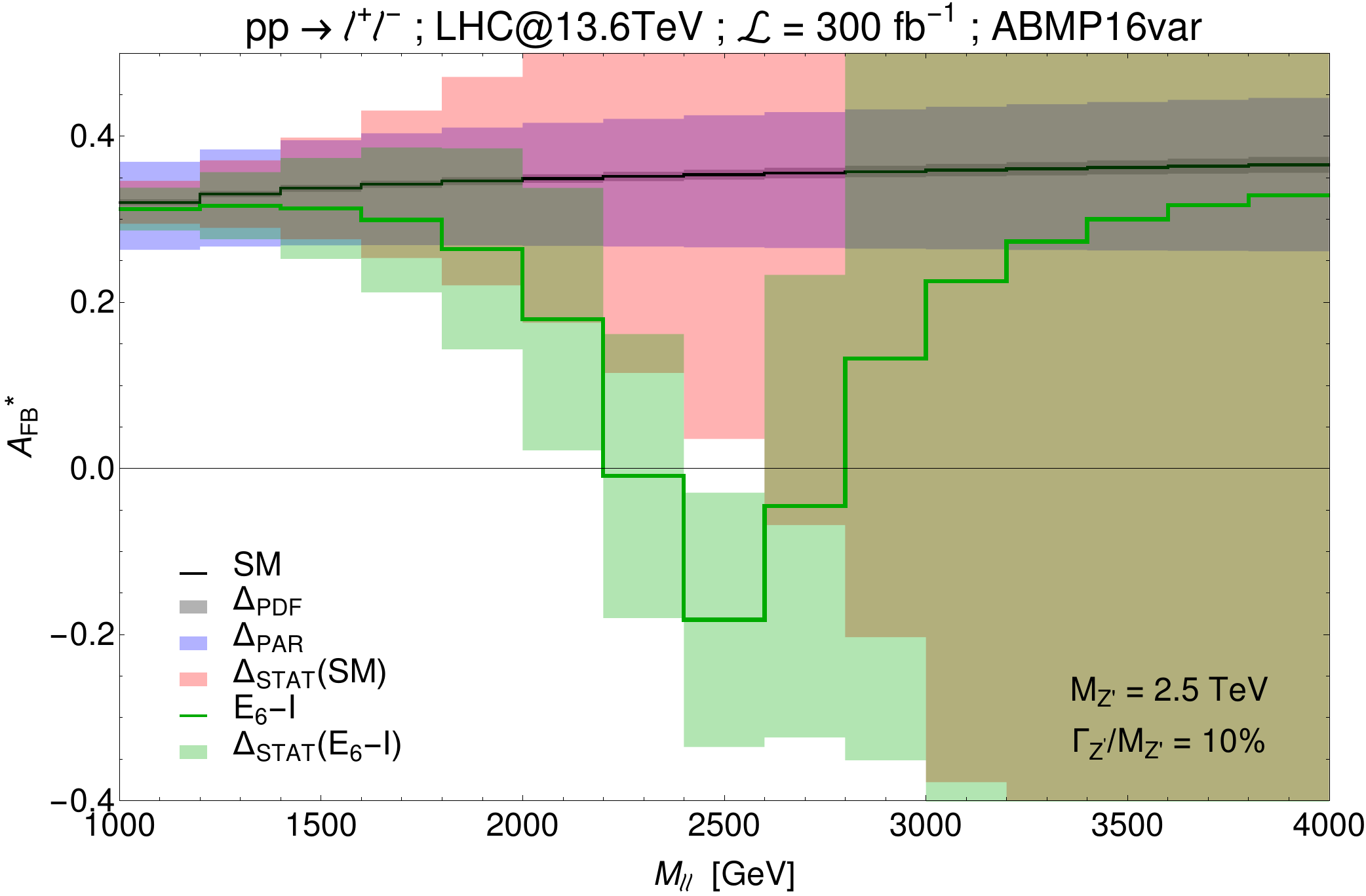}
\caption{Differential cross section (top) and $A_{\rm{FB}}^*$ (bottom) in invariant mass of the dilepton final state at the LHC with $\sqrt{s} = 13.6$ TeV for a $Z^\prime$ in the $E_6$-I model, where we set $M_{Z^\prime} = 2.5$ TeV and $\Gamma_{Z^\prime} / M_{Z^\prime} = $ 10\%.
The statistical uncertainty band corresponds to an integrated luminosity $\mathcal{L}$ = 300 fb$^{-1}$, combining the statistics of the di-electron and di-muon channels.}
\label{fig:bsm-benchmark}
\end{center}
\end{figure}

The signal profiles in the $A_{\rm{FB}}^*$ observable of the selected benchmark models are shown in Fig.~\ref{fig:afb-bsm} for the two configurations of narrow (top) and broad (bottom) resonances.
Despite the larger uncertainty band from the exponent variation method, the $A_{\rm{FB}}^*$ $Z^\prime$ signal shape remains well visible above the SM background predictions in both scenarios of narrow and broad resonances.
The property of the $A_{\rm{FB}}^*$ of being to some extent unaffected by variations of the resonance width is a crucial feature that makes this observable a suitable discriminant in BSM searches~\cite{Accomando:2015cfa}.

Next, in order to assess the sensitivity of the LHC to wide $Z^\prime$ resonances, we perform a statistical analysis over a selected $Z^\prime$ benchmark model.
We consider the $E_6$-I model~\cite{Hewett:1988xc} and set the $Z^\prime$ mass to 2.5 TeV and its width to 10\% of its mass.
In Fig.~\ref{fig:bsm-benchmark} we show the number of events (top) and the $A_{\rm{FB}}^*$ (bottom) with their statistical error in comparison with the systematic uncertainties from PDFs and from PDF parameterisation.
For a realistic estimation of the statistics of such a signal, we include Next-to-Next-to-Leading Order (NNLO) QCD corrections through a $K$-factor computed with 
\texttt{DYTurbo}~\cite{Camarda:2019zyx}, fixing the Electro-Weak (EW) parameters to the $G_\mu$ scheme at LO. The CMS experimental acceptances and efficiencies of the di-electron and di-muon channels~\cite{CMS:2021ctt} are included and the statistics of the two final states corresponding to an integrated luminosity of 300 fb$^{-1}$ are then combined.
In this scenario, it would be possible to disentangle the BSM signal from the background in both observables.
When considering only the statistical uncertainty, the significance could reach 4.4$\sigma$ and 4.3$\sigma$ for the bump search and for the $A_{\rm{FB}}^*$ observable, respectively.
On the other hand, once systematic uncertainties are included (linearly combining the two PDF errors while summing in quadrature the statistical uncertainties) the significance is reduced to 2.9$\sigma$ and 2.3$\sigma$ for the cross section and $A_{\rm{FB}}^*$ observable, respectively.
Therefore, in this context the significance from the $A_{\rm{FB}}^*$ still remains comparable to the cross section and, through the combination of the two, an earlier discovery can be achieved.

{\it Conclusion and outlook}.  Motivated by the 
observation~\cite{Fiaschi:2021sin} that  
 the LHC sensitivity to 
new gauge-boson broad resonances  is significantly enhanced 
owing to the quark PDF improvement 
from the DY lepton-charge and 
forward-backward asymmetry 
analysis~\cite{Fiaschi:2021okg}, in this paper we 
have  studied theoretical systematic
uncertainties in the multi-TeV mass region 
associated with the high-$x$ quark density. 
Based on a remark originally made 
in Ref.~\cite{Alekhin:2017kpj}, we have developed a method,   
which we term ABMP16var,  to take into 
account the physical 
effects of propagating, through QCD evolution, 
the low-scale non-perturbative parameterisation 
uncertainty in the falling-off $x \to 1$ quark 
density to the region of the very high mass tails in 
dilepton distributions. 
We have examined the implications of 
this method both on SM predictions for 
high-mass dilepton observables and 
 on BSM signals in a variety of 
narrow-$Z^\prime$ and wide-$Z^\prime$ models.

In the SM case, we find that the 
high-$x$ systematic uncertainties 
become extremely important in 
the multi-TeV mass region for both 
the cross section $ d \sigma / d M_{\ell \ell} $ and 
the asymmetry $A_{\rm{FB}}^*$, 
outweighing the 
uncertainties of standard PDF 
sets, and thus need to be 
taken into account for reliable 
estimates of theoretical errors on 
SM predictions in this region. Although 
they are larger than standard PDF 
uncertainties, we find that they do not 
spoil the positivity and near-flatness 
of $A_{\rm{FB}}^*$  at high masses which 
characterises the results from the 
PDF sets ABMP16, CT18 and MSHT20. 

On the other hand, we confirm the 
observation~\cite{Ball:2022qtp} on 
the $A_{\rm{FB}}^*$ result obtained from the 
PDF set NNPDF4.0  becoming steep 
for $ M_{\ell \ell} \gtap  4$ TeV,  
and eventually turning negative  
as a consequence of the high-sea, 
low-valence partonic content. It 
remains to be seen what size 
uncertainties the high-$x$ quark 
density systematics might give in 
the case of high-sea, low-valence 
NNPDF-like scenarios, and to what 
extent such scenarios are compatible with 
forward DY production measurements.  

Concerning BSM signals of  $Z^\prime$ states,   
we have selected benchmark models 
representative of the 
$E_6$, Generalised Left-Right  and Generalised Standard Model~\cite{Accomando:2010fz}, taking into account that 
narrow resonances with 
masses below 5 TeV are excluded by current 
 ATLAS and CMS 
bounds~\cite{ATLAS:2019erb,CMS:2021ctt}, while wide 
resonances are still allowed below 5 TeV. 
Taking $M_{Z^\prime} = 6$ TeV, 
$\Gamma_{Z^\prime} / M_{Z^\prime} = $ 1\% 
for a series of narrow-$Z^\prime$ 
models and 
$M_{Z^\prime} = 4$ TeV, $\Gamma_{Z^\prime} / M_{Z^\prime} = $ 20\%  for a series of wide-$Z^\prime$ 
models, we find that, even in the presence of 
systematic errors due to the high-$x$ quark density, 
computed via the ABMP16var method, 
the BSM signal profiles remain visible 
against the SM background in the 
3 - 6 TeV region of dilepton final states. This applies 
to both the cross section and $A_{\rm{FB}}^*$. 

We have further carried out a statistical analysis 
of LHC Run 3 sensitivity to wide $Z^\prime$ resonances, 
for integrated luminosity $\mathcal{L}$ = 300 fb$^{-1}$, 
using the $E_6$-I model with $M_{Z^\prime} = 2.5$ TeV 
and $\Gamma_{Z^\prime} / M_{Z^\prime} = $ 10\%.   
Evaluating the reduction in the significance due to 
the systematic uncertainties added to the 
statistical ones indicates that  
the significance from  $A_{\rm{FB}}^*$  
remains comparable to that of the cross section, and 
that discovery can be achieved by 
combining the two.   It is worth stressing that, 
going to higher invariant mass regions, one 
will have to deal with statistical limitations, particularly affecting $A_{\rm{FB}}^*$, at Run 3 of the LHC. Yet, many $Z^\prime$ models of the above kind would still be accessible herein.

Our analysis could well be extended to the HL-LHC, affording one  a tenfold increase in luminosity with respect to Run 3 of the LHC. However, in order to have a realistic estimate of the sensitivity of this LHC future configuration to the potential $Z^\prime$ signals considered here, one would need to make some assumptions about the final yield of the current stage of the CERN machine (i.e., either an increased limit on $M_{Z^\prime}$ or  a hypothesis of evidence or discovery for some values of it), project the PDF uncertainty following the fits to Run 3 data,  and face the possibility of (recall the bottom frame of Fig.~\ref{fig:afb-SM}) entering the 5 TeV (and above) mass region, where PDF predictions for the central value of $A_{\rm{FB}}^*$ vary greatly between different sets at present, and may not have been reconciled by then. Moreover, assuming that no anomalous dilepton events will have been seen in the relevant mass region at Run 3, so that higher $Z^\prime$ masses than those studied here will have to be considered at the HL-LHC, it is not certain that the increase in luminosity will offset the decrease in the $Z^\prime$ cross section (recall Fig.~\ref{fig:invmass}).
(Conversely, in the presence of a dilepton excess at Run 3, the HL-LHC would obviously serve the purpose of further $Z^\prime$ diagnostic.)  
For all such reasons, we  leave an HL-LHC analysis to future investigations.

\vskip 0.3cm 
\noindent 
{\it Acknowledgments}. 
The work of J.~Fiaschi has 
been supported by STFC under the Consolidated Grant ST/T000988/1.   
F.~Hautmann acknowledges funding from the 
Chinese Academy of Sciences President's International 
Fellowship Initiative, grant No.~2022VMA0005.
S.~Moretti is supported in part through the NExT Institute and acknowledges funding from the STFC Consolidated Grant ST/L000296/1.
S.~Moch is supported in part by the Bundesministerium f\"ur Bildung und Forschung under contract 05H21GUCCA.

\bibliography{refsdilept}



\end{document}